\documentclass[a4paper]{spie}  
\usepackage[]{graphicx}
\usepackage{units}
\usepackage{amsmath}
\title{Infinite impulse response modal filtering in visible adaptive optics} 

\author{G. Agapito\supit{a}, C. Arcidiacono\supit{b}, F. Quir\'os-Pacheco\supit{a}, A. Puglisi\supit{a}, S. Esposito\supit{a}
\skiplinehalf
\supit{a}Osservatorio Astrofisico di Arcetri, Largo E. Fermi 5, Firenze, Italy;\\
\supit{b}Osservatorio Astronomico di Bologna, Via Ranzani 1, Bologna, Italy;
}

\authorinfo{Further author information: (Send correspondence to G. Agapito)\\G. Agapito: E-mail: agapito@arcetri.astro.it} 

\begin{document}

\def\arcsec{$^{\prime\prime}$}
\def\araa{Annual Review of Astronomy and Astrophysics}
\def\apj{The Astrophysical Journal}
\maketitle 

\begin{abstract}
Diffraction limited resolution adaptive optics (AO) correction in visible wavelengths requires
a high performance control.
In this paper we investigate infinite impulse response filters that optimize the wavefront correction:
we tested these algorithms through full numerical simulations of a single-conjugate AO system comprising
an adaptive secondary mirror with 1127 actuators and a pyramid wavefront sensor (WFS).
The actual practicability of the algorithms depends on both robustness and knowledge of the real system:
errors in the system model may even worsen the performance.
In particular we checked the robustness of the algorithms in different conditions,
proving that the proposed method can reject both disturbance and calibration errors.
\end{abstract}

\keywords{Adaptive optics, Pyramid wavefront sensor, telescope vibrations, adaptive secondary mirror, modal control optimization}

\section{Introduction} \label{sec:intro}

Modern 8m telescopes are able to provide up to 90\% Strehl Ratio (SR) on high flux regime in
{\bf H} band (Esposito \emph{et al.}\cite{FLAO}), and reach a maximum resolution of about 30mas
(FLAO@LBT at the {\bf J} band or the Hubble Space Telescope at the {\bf U} band
or using lucky imaging technique on large telescope
at {\bf R} band Law \emph{et al.}\cite{2009ApJ...692..924L}).
Since the maximum achievable resolution is fixed
by diffraction retrieving Point Spread Function (PSF),
which Full Width at Half Maximum (FWHM) is $\lambda/D$, where $\lambda$
is the imaged wavelength and $D$ the telescope diameter.
8-m class telescope could improve the maximum achievable resolution
making images at shorter wavelenghts, such as {\bf V} band,
which could provide a 12mas resolution.
However behind these numbers is hidden a limitation:
an Adaptive Optics (AO) system for visible wavelengths
has tighter requirements compared to an near infra-red one.
In fact shorter wavelengths correspond to smaller Fried's parameters, $r_{0}$,
hence more degrees of correction and low measurement noise values
are required to obtain a good correction.\\
Optimized control techniques  such as a Modal Control Optimization\cite{GENDRON1994}, observer-based
and model-based\cite{Roux:04,kal,Kulcsar:06,Fedrigo:08,chellabi:09,Poyneer:10,agapito:11}
approaches can increase the performance of an AO system
in comparison to an integrator controller.
These methods need to work in realistic conditions,
\emph{i.e.} without an accurate knowledge of the real system and with complex disturbances
like calibration errors and telescope vibrations.
In particular, Modal Control Optimization has only one degree of freedom per mode, that is the integrator gain,
so it cannot efficiently reject disturbance condition such as telescope vibrations,
while the other methods rely on the accuracy of the system model.\\
For these reasons, starting from the work of Dessenne \emph{et al.}\cite{1998ApOpt..37.4623D},
we have chosen to design a method that is data driven, so it relies as little as possible
on the system model, and can have many degrees of freedoms as to adapt itself to the disturbances.\\
The remainder of this paper is as follows.
Sec.\ref{sec:cont} introduces the Infinite Impulse Response 
(IIR) filter based control along with other controllers that
we will use in the performance evaluation as comparison.
Sec.\ref{sec:etes} briefly describes the AO system considered in our simulations.
Finally in Sec.\ref{sec:sim} the controllers are analyzed through numerical simulations
from both performance and robustness viewpoint.

\section{AO Control Systems} \label{sec:cont}

\begin{figure}[ht]
	\begin{center}
		\includegraphics[width=0.75\textwidth]{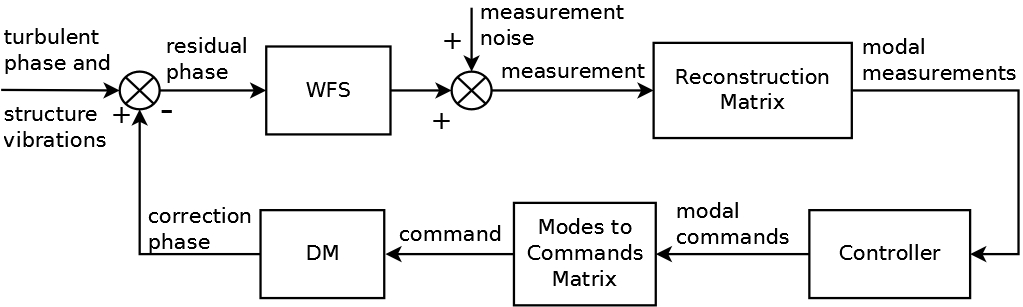}
	\end{center}
	\caption{Adaptive Optics Control loop.}
	\label{fig:aoloop}
\end{figure}
The AO system control loop considered in this work is represented in Fig. \ref{fig:aoloop}.
The controller gets the modal measurements as input and gives the modal commands as output.
The most used controller in AO systems is the integrator, so we have chosen it to make a comparison
that will be presented in the next sections.

\subsection{IIR filter based control} \label{sec:iir}

To determine the optimal control expressed as an IIR filter we realized an algorithm that minimized the
residual variance for each mode $i$ as a function of the IIR filter parameters:
\begin{equation}
	\label{eq:iir_min}
	[\hat{a}_i,\hat{b}_i] = \min_{[a_i,b_i]}J_i \, ,
\end{equation}
where $a_i$ and $b_i$ are two vectors of the IIR filter denominator and numerator coefficients:
\begin{equation}
	\label{eq:iir}
	C_i(z) = \frac{\sum\limits_{j=1}^{N_b}b_i(j) z^{-j}}{\sum\limits_{l=1}^{N_a}a_i(l) z^{-l}} \, ,
\end{equation}
where $N_a$ and $N_b$ are the order of respectively the denominator and numerator.
When the modes are orthonormal, minimizing the sum of the residual modal variances
is equivalent to minimizing the residual phase variance in the direction of the Natural Guide Star (NGS),
which is the goal of the control system studied in this paper.
The cost, that is the measured residual modal variance of the i-\emph{th} mode $J_i$, is:
\begin{equation}
	\label{eq:iir_cost}
	J_i = \sum\limits_{f=\frac{T}{n}}^{\frac{T}{2}} \Phi_{i}^{meas}(\omega)= \sum\limits_{f=\frac{T}{n}}^{\frac{T}{2}} \| W(z) \|^2 \Phi_{i}^{ol}(\omega) \, ,
\end{equation}
where:
\begin{equation}
	\label{eq:W}
	W(z) =  \frac{1}{1 + \mathrm{C}(z) \mathrm{WFS}(z) \mathrm{DM}(z)}.
\end{equation}
$\omega$ is the discrete frequency, $n$ is the number of points of the discrete signals,
$T$ is the sampling period, $\Phi_{i}^{meas}(\omega)$ is the Power Spectral Density (PSD)
of the closed loop i-\emph{th} modal coefficient determined from the WFS measurements, and
$\Phi_{i}^{ol}(\omega)$ is the PSD of the open loop i-\emph{th} modal coefficient.
The open loop measurement $o_i$ would be the input signal of the AO system without the control feedback,
and we reconstruct it as the sum of the closed loop measurement $s_i$ and command $c_i$. At iteration $k$,
taking into account the delay $d$, it can be written as:
\begin{equation}
	\label{eq:ol}
	o_i(k) = s_i(k) + c_i(k-d) \, .
\end{equation}
It can be shown (see Dessenne \emph{et al.}\cite{1998ApOpt..37.4623D}) that:
\begin{equation}
	\label{eq:ol_cl}
	s_i(k) = W(z) o_i(k) \, ,
\end{equation}
so that the relationship shown in Eq.\ref{eq:iir_cost} is demonstrated.\\
This algorithm must verify the stability of the loop during the minimization of the cost $J_i$,
and it must discard the controller that do not stabilize the loop.
In fact, an unstable controller could produce the minimum cost, but it could not be implemented.\\
To take into account the modelling errors, like pupil shifts or approximations of DM and WFS models,
a stability constrain is not enough.
Therefore the controller must guarantee some robustness,
so small variations of the model parameters would not drive the controller to instability.
Thus we impose $\max \| {\mu} \left(  \chi_i(z) \right) \|^2 \le \eta$, where $\mu(\cdot)$
gives the roots of the polynomial, $0<\eta<1$, and $\chi_i(z) = 1 + C_i(z) P(z)$
is the characteristic polynomial, where $P(z)$ is the AO system model:
\begin{equation}
	\label{eq:plant}
	P(z) = \mathrm{WFS}(z) \mathrm{DM}(z) \, ,
\end{equation}
we have simply assumed that $\mathrm{WFS}(z) = z^{-1}$ and $\mathrm{DM}(z) = z^{-1}$,
in order to model a simple AO loop with two frames delay.\\
Note that $\eta$ gives a measurement of the stability margin since
if $\eta$ is equal to 1 only stability is guaranteed,
while smaller values of $\eta$ correspond to a greater distance from instability,
\emph{i.e.} a greater robustness.\\
In this work we choose $\eta=0.8$, which corresponds to a gain margin of 0.2.
Hence the algorithm can be summarized as:
\begin{itemize}
\item Acquire with a closed loop the mode measurement $s_i$ and commands $c_i$.
\item Determine $\Phi_{i}^{ol}(f)$.
\item Choose $N_a$ and $N_b$ and the starting IIR filter parameters $a_i(0)$, $b_i(0)$.
\item Search the combination of  IIR filter parameters $\hat{a}_i$, $\hat{b}_i$ which minimizes $J_i$ and satisfies $\max \| {\mu} \left(  \chi_i(z) \right) \|^2 \le \eta$.
\end{itemize}
In this work, the minimization of $J_i$ is based on the downhill
simplex method of Nelder \& Mead~\cite{NelderMead65}, and
we opted for $N_a = N_b = 3$ in case of turbulence
and 2 coefficients more for each vibration.
So, in the next simulations, Tip and Tilt have $N_a = N_b = 7$ while the other modes have $N_a = N_b = 3$,
because, as presented in Sec.\ref{sec:etes}, Tip and Tilt are affected by two vibrations.

\subsection{Modal Gain Integrator} \label{sec:int}

\begin{figure}[ht]
	\begin{center}
		\begin{minipage}[b]{0.45\linewidth}
			\includegraphics[width=\textwidth]{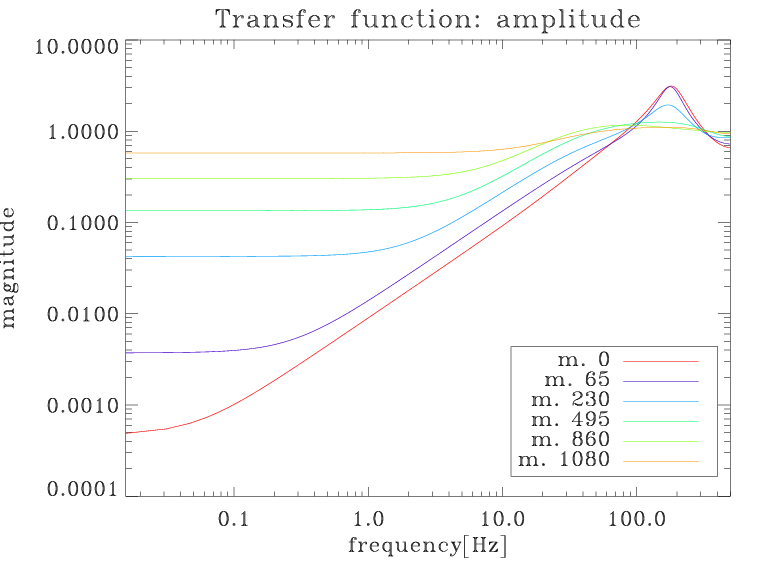}
		\end{minipage}
		\begin{minipage}[b]{0.45\linewidth}
			\includegraphics[width=\textwidth]{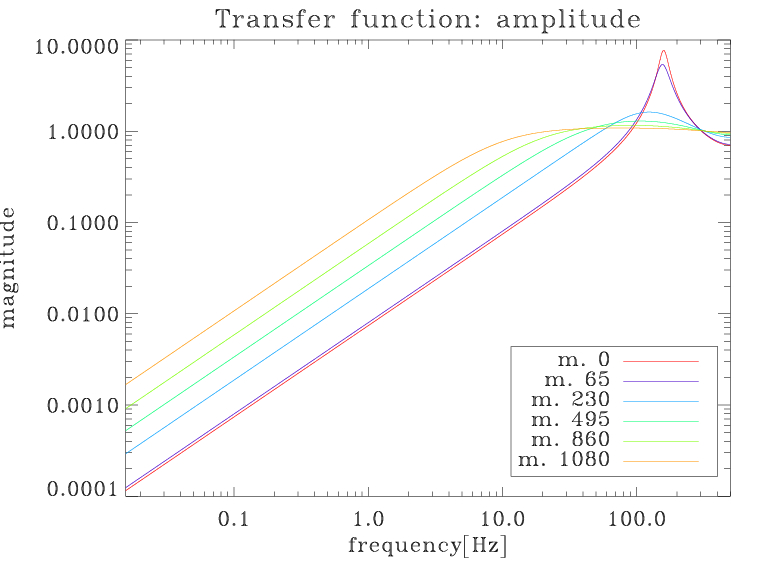}
		\end{minipage}
	\end{center}
	\caption{RTFs for various modes with the IIR filter based control with $N_a = N_b = 3$ (left), and the OMGI (right). This RTFs are determined for the case magnitude 12.5 without vibrations nor pupil shifts. All the other parameters may be found in Tab.\ref{tab:simparams}.}
	\label{fig:tfm}
\end{figure}

The Transfer Function of the modal integrator is:
\begin{equation}
	\label{eq:int}
	C(z) = \frac{g_i}{1-z^{-1}}
\end{equation}
where $g_i$ is the integrator gain of the \emph{i}-th mode, and $z$ is the variable of the Z-transform.\\
In this work, we refer as integrator the particular case in which all the gains are equal,
and as Optimized Modal Gain Integrator\cite{GENDRON1994} (OMGI) the particular case
in which each modal gain is optimized.
For both cases we optimize the gain: in the integrator case we optimized the global gain
by a trial and error procedure \emph{i.e.} we ran many simulations with different gains and
then we chose the one that produced the best results in terms of Strehl Ratio (SR);
instead, in case of OMGI, we determine the gain following the same algorithm presented in the previous section,
but instead of minimizing the cost $J_i$ as a function of the IIR filter parameters,
we minimized it as function of the integrator gain $g_i$:
\begin{equation}
	\label{eq:omgi_min}
	\hat{g}_i = \min_{g_i}J_i \, ,
\end{equation}
since the integrator controller is an IIR filter with $N_a=2$ $N_b=1$ and a single variable parameter,
the gain $g$.\\
An example of IIR and OMGI Rejection Transfer Functions (RTF) $W$ is shown in Fig.\ref{fig:tfm}.
Note that the OMGI give better rejection at the lowest frequencies, but the bandwidth
rapidly decreases with the mode order.
The integrator controller bandwidth is in direct proportion to the peak at high frequencies,
and this peak increase the residual phase due to the measurement noise.
Hence, the high order modes, that are more affected by noise, have a lower bandwidth.
Instead the IIR filter gives a bandwidth greater than $\unit[20]{Hz}$ for all the modes
with low peaks at high frequencies.

\section{AO system and simulator} \label{sec:etes}

The AO system analyzed in this work is based on a pyramid wavefront sensor (WFS)
with tilt modulation (Ragazzoni\cite{pyramid}) and on an adaptive secondary mirror (ASM)
(Salinari \emph{et al.} \cite{Salinari94}).
We chose to work with an 8m class telescope, an ASM with 1127 actuators,
and a WFS with 40 $\times$ 40 sub-apertures.
As detector of the WFS we considered an ANDOR iXon X3 897 camera\cite{ANDOR:EMCCD}
with Read Out Noise (RON), Excess Noise, and Clock Induced Charge (CIC) (Daigle \emph{et al.}\cite{Daigle:04})
values shown in Tab.\ref{tab:simparams}.
All simulations reported hereafter rely on the End-to-End numerical simulator used to perform
the performance analysis and optimization of the First Light Adaptive Optics (FLAO)
system of the LBT telescope~\cite{2010SPIE.7736E.116Q}.\\
The atmospheric turbulence was simulated with a set of phase screens with Von-Karman statistics that were
displaced in front of the telescope pupil to emulate the time-evolving turbulence according to the Taylor
hypothesis.\\
Pupil shifts due to the sensor optical misalignments are a common calibration error.
We will evaluate in our simulations the robustness of the controller to this kind of errors.
The pupil shifts considered in the simulations have an amplitude of $\unit[0.1]{pixel}$ in both
\emph{x} and \emph{y} directions.\\
We add in the simulations, as external disturbance, telescope vibrations.
Vibrations may arise from many different situations, e.g. telescope
orientation, telescope tracking errors, and wind shaking.
In particular, since vibrations cause displacements of the image
they have a major impact on so-called tip/tilt modes.
Hence, we chose to introduce Tip/Tilt vibrations with central frequencies of $\unit[13]{Hz}$ and $\unit[22]{Hz}$,
an RMS of $\unit[20]{mas}$ each,
and a damping ratio of $0.01$.
\begin{table}[ht]
	\begin{center}
		\begin{minipage}[b]{0.52\linewidth}
			\begin{tabular}{|l|c|}
				\hline
				\multicolumn{2}{|c|} {\textbf{Telescope}}\\
				\hline
				Effective diameter ($D$) & $\unit[8]{m}$ \\
				\hline
				Central obstruction & $\unit[0.138]{D}$ \\
				\hline
				\hline
				\multicolumn{2}{|c|}{\textbf{Pyramid WFS}}\\
				\hline
				Sensing wavelength ($\lambda$) & $\unit[0.75]{\mu m}$\\
				\hline
				Number of sub-apertures & $40 \times 40$\\
				\hline
				Tilt modulation radius & $2.0\frac{\lambda}{D}$\\
				\hline
				Total average transmission & $0.41$\\
				\hline
				RON (electrons per pixel) & $\unit[0.06]{e^{-}RMS}$\\
				\hline
				Clock Induced Charge & $\unit[0.005]{e^{-}/pixel/frame}$\\
				\hline
				Excess Noise & $\sqrt{2}$\\
				\hline
				\hline
				\multicolumn{2}{|c|}{\textbf{Guide star}}\\
				\hline
				Mag. zero point (Johnson~\cite{1966ARAA...4..193J}) & $\unit[1.76 \, 10^{-8}]{J/s/m^2/\mu m}$ \\
				\hline
			\end{tabular}
		\end{minipage}
		\begin{minipage}[b]{0.38\linewidth}
			\begin{tabular}{|l|c|}
				\hline
				\multicolumn{2}{|c|} {\textbf{ASM}}\\
				\hline
				Modes (Karhunen-Lo\`eve) & $1127$ \\ 
				\hline
				\hline
				\multicolumn{2}{|c|}{\textbf{Turbulence}}\\
				\hline
				Outer scale ($L_0$) & $\unit[25]{m}$\\
				\hline
				Mean wind speed & $\unit[10.6]{m/s}$\\
				\hline
				\hline
				\multicolumn{2}{|c|}{\textbf{Telescope Vibration}}\\
				\hline
				Frequencies & $13, \unit[22]{Hz}$\\
				\hline
				Standard deviation & $\unit[20]{mas}$\\
				\hline
				Damping ratio & $0.01$\\
				\hline
				\hline
				\multicolumn{2}{|c|}{\textbf{Loop paramenters}}\\
				\hline
				Sampling frequency & $\unit[1000]{Hz}$\\
				\hline
				Delay & 2 frames\\
				\hline
			\end{tabular}
		\end{minipage}
		\caption{Summary of the simulation parameters.}
		\label{tab:simparams}
	\end{center}
\end{table}

\section{Simulations} \label{sec:sim}

\begin{figure}[ht]
	\begin{center}
			\includegraphics[width=0.5\textwidth]{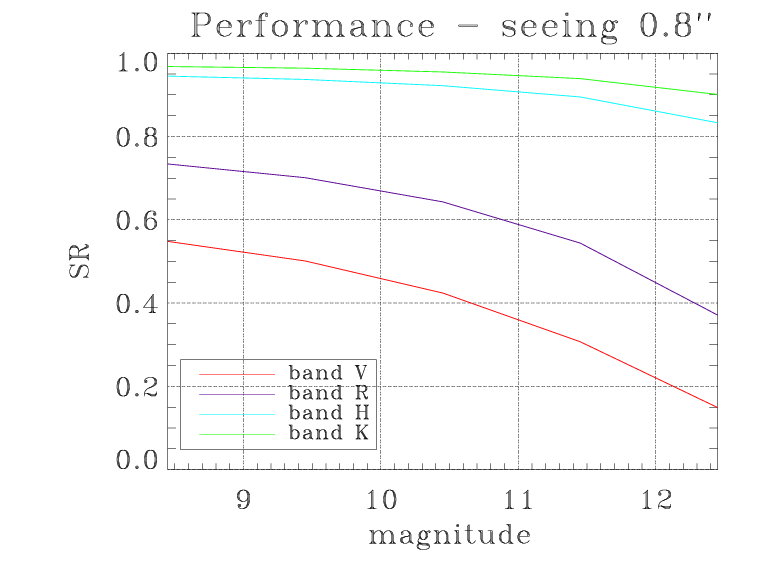}
	\end{center}
	\caption{Summary of simulation results for the integrator controller in case of turbulence without pupil shifts nor vibrations. Integrator gain is determined as described in Sec.\ref{sec:int}. Seeing 0.8\arcsec, see Table~\ref{tab:simparams} for the other parameters.}
	\label{fig:sim08}
\end{figure}

\begin{figure}[ht]
	\begin{center}
		\begin{minipage}[b]{0.45\linewidth}
			\includegraphics[width=\textwidth]{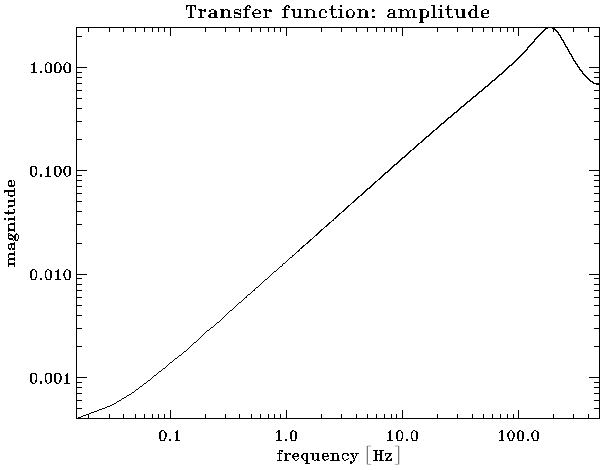}
		\end{minipage}
		\begin{minipage}[b]{0.45\linewidth}
			\includegraphics[width=\textwidth]{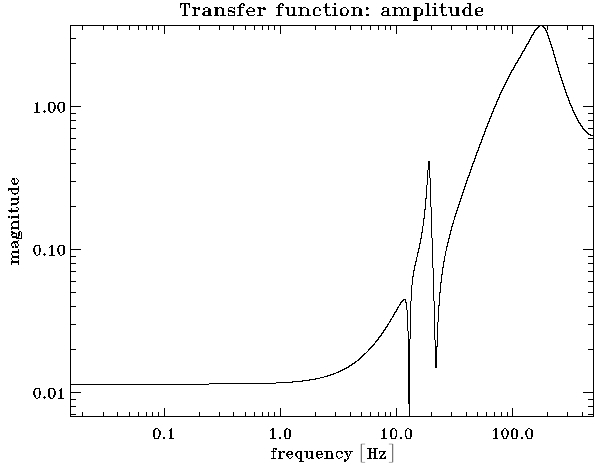}
		\end{minipage}
	\end{center}
	\caption{Rejection Transfer Function (RTF) for the Tip mode with an IIR filter based control with $N_a = N_b = 3$ (left), and with $N_a = N_b = 7$ (right). Note that the order 7 is used to reject both vibrations at $\unit[13]{}$ and $\unit[22]{Hz}$ and atmospheric turbulence. These RTFs are determined for the case magnitude 8.5. All the other parameters may be found in Tab.\ref{tab:simparams}}
	\label{fig:iirtv}
\end{figure}

\begin{figure}[ht]
	\begin{center}
		\begin{minipage}[b]{0.4\linewidth}
			\includegraphics[width=\textwidth]{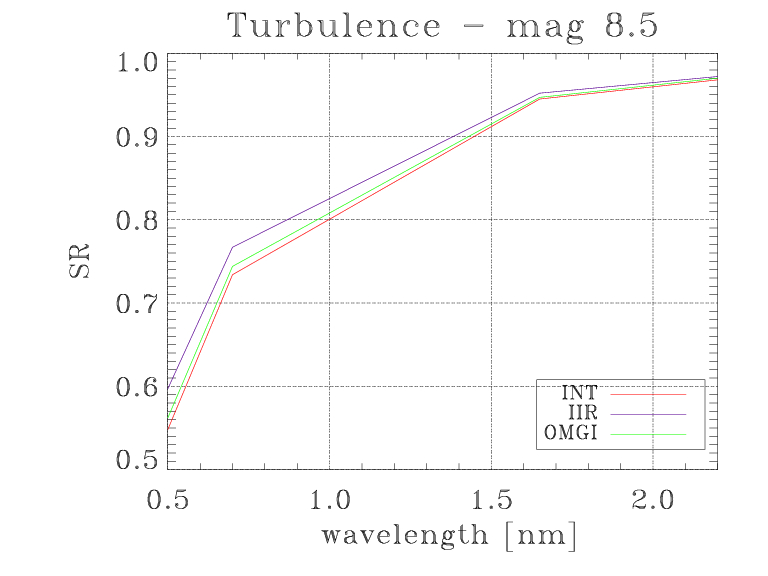}
		\end{minipage}
		\begin{minipage}[b]{0.4\linewidth}
			\includegraphics[width=\textwidth]{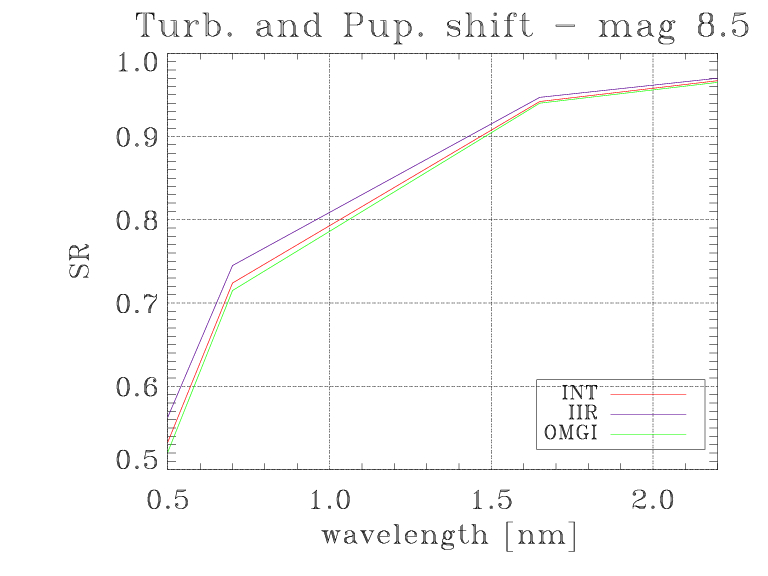}
		\end{minipage}
		\begin{minipage}[b]{0.4\linewidth}
			\includegraphics[width=\textwidth]{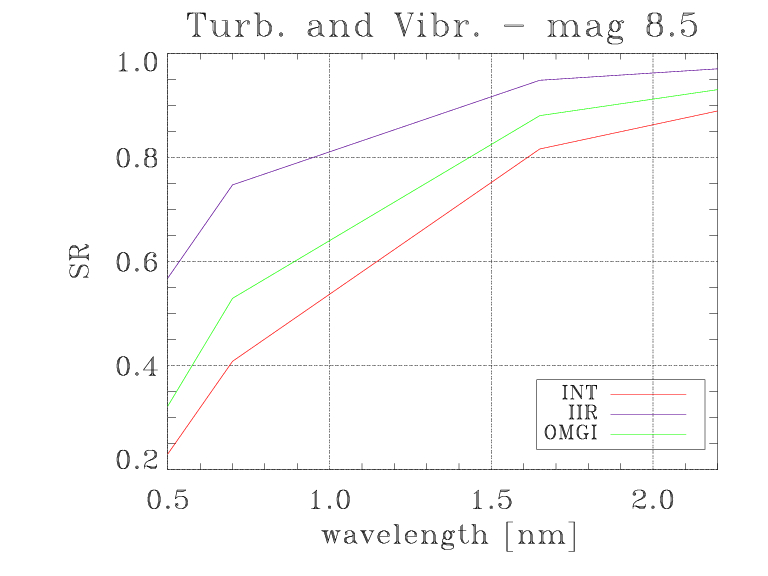}
		\end{minipage}
		\begin{minipage}[b]{0.4\linewidth}
			\includegraphics[width=\textwidth]{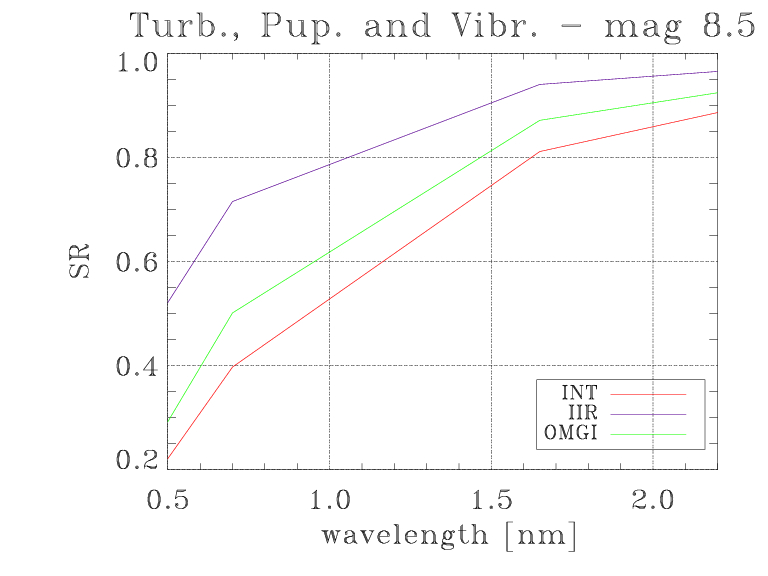}
		\end{minipage}
		\end{center}
		\caption{Summary of simulation results for turbulence, pupil shifts and vibrations at magnitude 8.5. Comparison between Integrator, IIR filter based control, and autogain. Seeing 0.8\arcsec.}
		\label{fig:simINTIIRAUT8}
\end{figure}

\begin{figure}[ht]
	\begin{center}
		\begin{minipage}[b]{0.4\linewidth}
			\includegraphics[width=\textwidth]{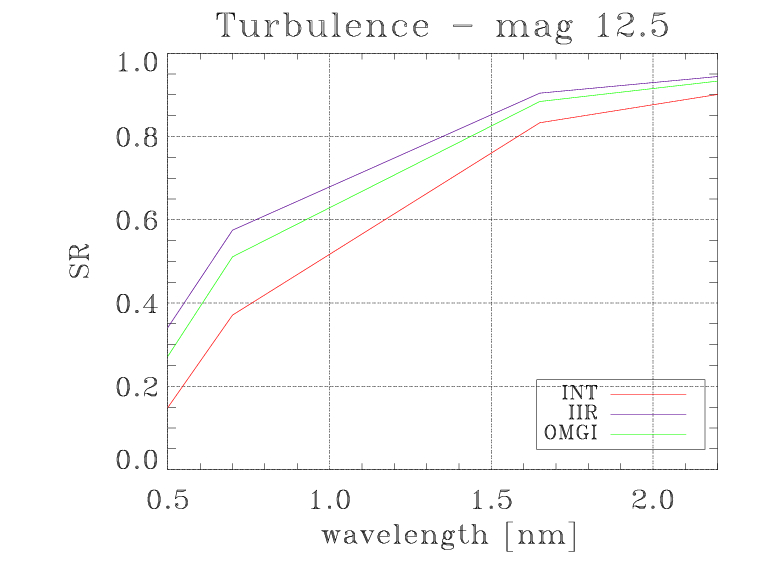}
		\end{minipage}
		\begin{minipage}[b]{0.4\linewidth}
			\includegraphics[width=\textwidth]{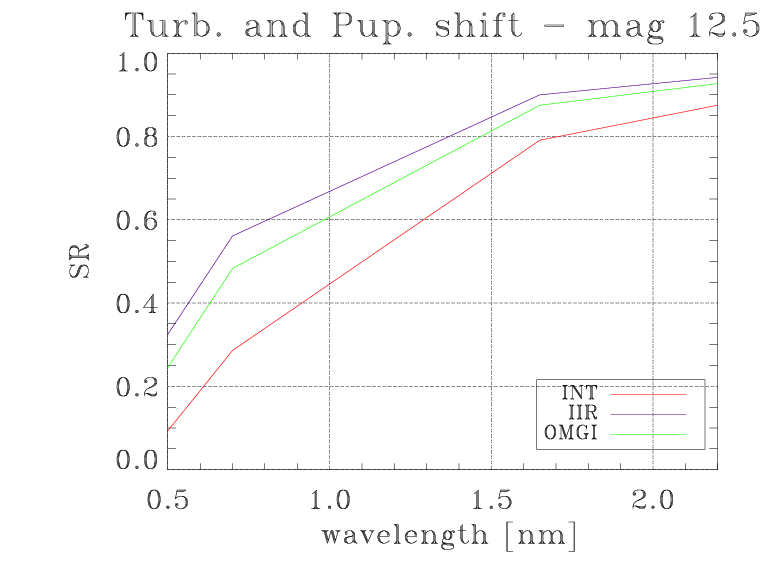}
		\end{minipage}
		\begin{minipage}[b]{0.4\linewidth}
			\includegraphics[width=\textwidth]{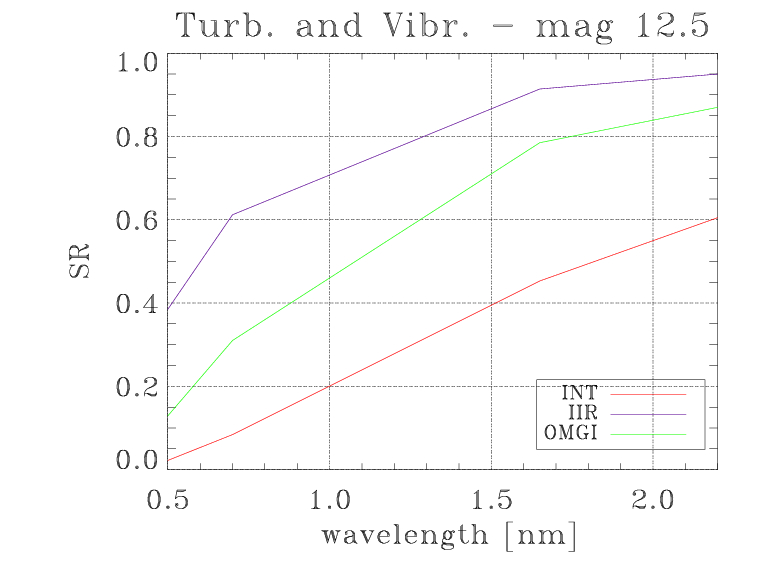}
		\end{minipage}
		\begin{minipage}[b]{0.4\linewidth}
			\includegraphics[width=\textwidth]{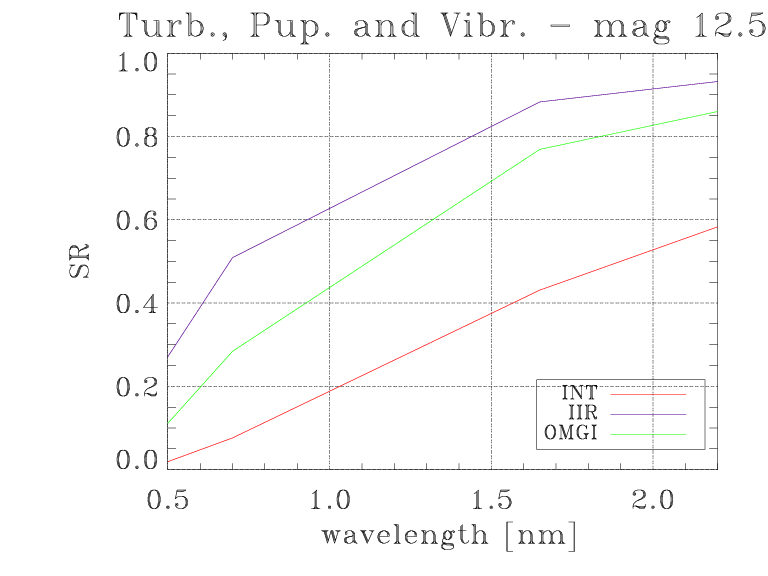}
		\end{minipage}
	\end{center}
	\caption{Summary of simulation results for turbulence, pupil shifts and vibrations at magnitude 12.5. Comparison between Integrator, IIR filter based control, and autogain. Seeing 0.8\arcsec.}
	\label{fig:simINTIIRAUT12}
\end{figure}

Firstly we will check the performance of the integrator with turbulence
and without pupil shifts nor vibrations.
We chose to run all the simulations in between magnitude 8.5 and 12.5.
All the simulation parameters are listed in Tab.\ref{tab:simparams}.
The gain is determined with the method described in Sec.\ref{sec:int}.
In Fig.\ref{fig:sim08} and Tab.\ref{tab:simres} are shown the results.
As we expect the performance of the system rapidly decreases at the shortest wavelengths.
Moreover the decrease is greater in the visible wavelengths,
in fact, while at {\bf K} band from magnitude $8.5$ to $12.5$
less than $8\%$ of SR is lost, at {\bf V} band the difference is about $40\%$.\\
\begin{table}[ht]
	\begin{center}
		\begin{tabular}{|l||c|c|c|c|}
			\hline
			band & V & R & H & K \\
			\hline
			magnitude & \multicolumn{4}{|c|} {\textbf{Strehl Ratio} $\%$ } \\
			\hline
			 8.5 &   54.8 &   73.4 &   94.5 &   96.8\\ \hline
			 9.5 &   50.1 &   70.1 &   93.7 &   96.4\\ \hline
			10.5 &   42.4 &   64.3 &   92.2 &   95.5\\ \hline
			11.5 &   30.7 &   54.4 &   89.5 &   93.9\\ \hline
			12.5 &   14.9 &   37.1 &   83.3 &   90.1\\ \hline
		\end{tabular}
		\caption{Summary of simulation results for the integrator controller. Integrator gain is determined as described in Sec.\ref{sec:int}. Seeing 0.8\arcsec, see Table~\ref{tab:simparams} for the other parameters.}
		\label{tab:simres}
	\end{center}
\end{table}
In the next simulations we considered only these two magnitudes, $8.5$ and $12.5$,
to make a comparison between all the controllers.
We will compare the controller in 4 difference observation conditions:
the first one is with turbulence and without pupil shifts nor vibrations,
then second and third case with turbulence and with pupil shifts or vibrations,
respectively, and the last one with all the disturbances included.\\
The controllers parameters are optimized for each condition.
The gain found for the integrator and the minimum, maximum and mean gain of the OMGI
are shown in Tab.\ref{tab:simGAIN}.
Regarding the IIR filter Fig.\ref{fig:iirtv} shown, as an example, two RTFs
for the tip/tilt modes in two different cases:
on the left, with only turbulence and,
on the right, with turbulence, pupil shifts, and vibrations.
Note that in the first case we design a filter with $N_{a}=N_{b}=3$,
while with all the disturbances $N_{a}=N_{b}=7$.\\
\begin{table}[ht]
	\begin{center}
		\begin{tabular}{|l||c||c|c|c|}
			\hline
			Controller &  integrator &  \multicolumn{3}{|c|}{OMGI} \\
			\hline
			magnitude & g & g min & g max & g mean\\
			\hline
			\hline
			\multicolumn{5}{|c|}{\textbf{Turbulence}} \\
			\hline
			 8.5 &  0.6 & 0.23 & 0.88 & 0.50 \\ \hline
			12.5 &  0.3 & 0.01 & 0.85 & 0.20 \\ \hline
			\hline
			\multicolumn{5}{|c|}{\textbf{Turbulence + Pupil shifts}} \\
			\hline
			 8.5 &  0.6 & 0.08 & 0.92 & 0.33 \\ \hline
			12.5 &  0.3 & 0.03 & 0.85 & 0.23\\ \hline
			\hline
			\multicolumn{5}{|c|}{\textbf{Turbulence + Vibrations}} \\
			\hline
			 8.5 &  0.8 & 0.23 & 0.92 & 0.50 \\ \hline
			12.5 &  0.4 & 0.01 & 0.93 & 0.22 \\ \hline
			\hline
			\multicolumn{5}{|c|}{\textbf{Turbulence + Pupil shifts + Vibration}} \\
			\hline
			\hline
			 8.5 &  0.7 & 0.08 & 0.92 & 0.33 \\ \hline
			12.5 &  0.4 & 0.01 & 0.92 & 0.23 \\ \hline
		\end{tabular}
		\caption{Summary of simulation optimal gain. Comparison between Integrator and OMGI. Seeing 0.8\arcsec.}
		\label{tab:simGAIN}
	\end{center}
\end{table}
All the results of the simulations are summarized in Tab.\ref{tab:simINTIIRAUT} in term of SR
for all the considered observing conditions.
Fig.\ref{fig:simINTIIRAUT8} shows the SR at different wavelengths at magnitude 8.5, Fig.\ref{fig:simINTIIRAUT12}
shows the corresponding results at magnitude 12.5.\\
\begin{table}[ht]
	\begin{center}
		\begin{tabular}{|l||c|c|c|c||c|c|c|c||c|c|c|c|}
			\hline
			band & V & R & H & K & V & R & H & K & V & R & H & K \\
			\hline
			Controller &  \multicolumn{4}{|c||}{integrator} & \multicolumn{4}{|c||}{IIR filter} &  \multicolumn{4}{|c|}{OMGI} \\
			\hline
			magnitude & \multicolumn{12}{|c|}  {\textbf{Strehl Ratio} $\%$ } \\
			\hline
			\multicolumn{13}{|c|}{\textbf{Turbulence}} \\
			\hline
			 8.5 & 54.8 &   73.4 &   94.5 &   96.8 &   59.7 &   76.7 &   95.2 &   97.2 &   56.2 &   74.4 &   94.7 &   97.0 \\ \hline
			12.5 &  14.9 &   37.1 &   83.3 &   90.1 &   34.1 &   57.5 &   90.4 &   94.4 &   27.2 &   51.1 &   88.4 &   93.3 \\ \hline
			\hline
			\multicolumn{13}{|c|}{\textbf{Turbulence + Pupil shifts}} \\
			\hline
			 8.5 & 53.3 &   72.4 &   94.2 &   96.7 &   56.3 &   74.5 &   94.7 &   97.0 &   52.1 &   71.5 &   94.0 &   96.5 \\ \hline
			12.5 &  9.3 &   28.6 &   79.1 &   87.5 &   32.5 &   56.1 &   90.0 &   94.2 &   24.5 &   48.3 &   87.5 &   92.7 \\ \hline
			\hline
			\multicolumn{13}{|c|}{\textbf{Turbulence + Vibrations}} \\
			\hline
			 8.5 & 23.0 &   40.8 &   81.6 &   88.9 &   56.8 &   74.7 &   94.8 &   97.0 &   32.2 &   52.9 &   88.0 &   93.0 \\ \hline
			12.5 &  2.2 &    8.4 &   45.3 &   60.5 &   38.5 &   61.2 &   91.4 &   95.0 &   12.9 &   31.0 &   78.5 &   87.0 \\ \hline
			\hline
			\multicolumn{13}{|c|}{\textbf{Turbulence + Pupil shifts + Vibrations}} \\
			\hline
			\hline
			 8.5 &  22.1 &   39.7 &   81.1 &   88.6 &   52.1 &   71.5 &   94.0 &   96.5 &   29.2 &   50.1 &   87.1 &   92.4 \\ \hline
			12.5 &   1.9 &    7.6 &   43.1 &   58.3 &   27.1 &   50.9 &   88.3 &   93.2 &   11.1 &   28.4 &   76.9 &   86.0\\ \hline
		\end{tabular}
		\caption{Summary of simulation results. Comparison between Integrator, IIR filter based control, and OMGI. Seeing 0.8\arcsec.}
		\label{tab:simINTIIRAUT}
	\end{center}
\end{table}
The integrator and the OMGI show similar results in case of turbulence and pupil shifts,
while the vibrations affect more the integrator.
These results is expected, because the OMGI has more degrees of freedoms, one for each mode,
with respect to the single degree of the integrator.
In fact it can be shown in Tab.\ref{tab:simGAIN} that
the OMGI maximum gain value is greater in case of vibrations,
so it has a greater bandwidth and can better reject the Tip/Tilt disturbances.
Note that in case of pupil shifts the gains are lower
to guarantee greater robustness.
At the faintest magnitude the differences between the controller are greater,
because, as we expected, the IIR filters reject better the noise
without decreasing the bandwidth (see Sec.\ref{sec:int} and Fig.\ref{fig:tfm}).\\
The IIR filter based control produces similar performance in all the tested conditions:
it can efficiently reject both calibration errors and disturbances.
In the worst case -- turbulence, pupil shifts and vibrations -- the IIR filter
at {\bf V} band gives the same SR of the OMGI and
about two times the one of the integrator both in the condition
without pupil shifts nor vibrations.\\

\section{Processing power estimate}\label{sec:power}

In this section we compare the processing power required for the IIR filter based control described above, compared to the integrator case. Referencing the scheme in Fig.\ref{fig:aoloop}, we can split the required processing power into three major steps:
\begin{enumerate}
\item Reconstruction matrix \label{item:1}
\item Controller \label{item:2}
\item Modes to command projection matrix \label{item:3}
\end{enumerate}
Steps \ref{item:1} and \ref{item:3} are common to the different control strategies, while step \ref{item:2} is where the techniques differ.\\
For the estimate, we suppose the following system parameters (partially following Tab.\ref{tab:simparams}):
\begin{itemize}
\item Number of controlled modes: 1127 (including tip-tilt)
\item Number of DM actuators: 1127
\item Number of slopes measured by the WFS: 2480
\item Loop frequency: 1000 Hz
\item System delay: 2 frames
\end{itemize}
Step \ref{item:1} is a straightforward vector/matrix multiplication. The required processing power is $N_{\mathrm{modes}}\times N_{\mathrm{slopes}}$ multiply-accumulate (MAC, a basic operation employed by most Digital Signal Processors - DSPs where two operands are multiplied and summed to an accumulator). In our example it is about 2.8 Mega-MAC (MMAC). Step 3 is also a vector/matrix multiplication, this time of dimension $N_ {\mathrm{modes}}\times N_{\mathrm{commands}}$. In our example this amounts to about 1.3 MMAC. Assuming a total delay of 2 frames at 1 KHz, and further assuming that this computation cannot take more than 1/10th of the available time in order to be considered negligible, the processing power requirements for steps \ref{item:1} and \ref{item:3} combined is 20 GMAC/sec. This is comparable with existing systems, like the LBT one\cite{2003SPIE.5169..159R} where the total processing power, distributed over 168 DSPs, is about 25 GMAC/sec (only a fraction of this processing power is used on the LBT system, since the reconstrucion matrices have smaller dimensions and the DSPs also have more tasks to attend to). Both steps \ref{item:1} and \ref{item:3} can be efficiently parallelized over such a large number of DSPs, since each row of the matrix multiplication can be processed independently of the others.\\
For step \ref{item:2}, the integrator case needs to sum the measurement vector from the previous frame to the current one. Additionally, a gain value must be multiplied to each measurement value. This requires about 1 Kilo-MAC (KMAC), which is completely negligible (about 1/4000th) when compared to the previous numbers. The only effect (and a rather large one) is to prevent full pipelining of the two matrix multiplications.\\
For the IIR filters, we take into account a filter whith order 7 for the tip-tilt modes, and 3 for the other modes (as in Sec.\ref{sec:iir}). With $N_a=7$ and $N_b=7$, this requires only a few dozens MACs for the tip-tilt IIR filters, and about 6 KMAC for the other modes. While this number is 6 times bigger than the integrator, it is again negligible in the context of the processing power required by steps \ref{item:1} and \ref{item:3}. Furthermore, no additional requirements are posed on the other processing steps, except for the full separation of the two matrix multiplications, which was already required for the integrator with modal gain.\\
We can therefore conclude that the IIR filtering strategy outlined in this article can readily be implemented on existing or future AO systems with little to no penalty in terms of loop delay.

\section{Conclusions}

In this paper we have shown that IIR filter based control is a control solution
that can deliver good performance even in difficult conditions.
It is a possible choice for AO systems requiring
high performance and robustness in case of calibration errors
and telescope vibrations.\\
Moreover, thanks to the data driven approach, the presented method need only a basic knowledge of the system,
and the computational burden is focused in the parameters optimization that can be made off-line.
In fact, the computational power requested to the Real Time Computer of the AO system will be negligible
in comparison to the integrator as we proved in Sec.\ref{sec:power}.\\
As further work, more investigation on the robustness, on the coefficient $\eta$
and on the impact of IIR filter order $N_a$, $N_b$ on the performance
must be done.
\begin{acknowledgements}
	This study was supported by the TECNO INAF 2009 grant from the italian Istituto Nazionale di Astrofisica (INAF).
\end{acknowledgements}

\bibliography{biblio}
\bibliographystyle{spiebib}

\end{document}